\documentstyle[12pt]{article}
\renewcommand{\section}[1]{\refstepcounter{section}
\vspace{24pt}\noindent{\bf\arabic{section}.\quad #1}
\vspace*{12pt} \setcounter{subsection}{0}}
\renewcommand{\subsection}[1]{\refstepcounter{subsection}
\vspace{12pt}\noindent{\bf\arabic{section}.\arabic{subsection}.\quad #1}
\vspace*{6pt}}
\newcommand{\ulsect}[1]{\vspace{18pt}\noindent{\bf #1}
\vspace*{12pt}}

\begin{document}

\begin{flushright} CERN-TH.7038/93\\
\end{flushright}
\vspace*{10mm}
\begin{center}
{\bf Thermal quark production in ultra-relativistic nuclear
collisions}\\[10mm]
Tanguy Altherr$^*$ and David Seibert$^{\dag}$ \\[5mm] Theory
Division, CERN, CH-1211 Geneva 23, Switzerland\\[10mm]
{\bf Abstract}\\
\end{center}
\hspace*{12pt}
We calculate thermal production of $u$, $d$, $s$, $c$ and $b$ quarks in
ultra-relativistic heavy ion
collisions. The following processes are taken into account: thermal gluon
decay ($g\to i\overline{\imath}$),
gluon fusion ($gg\to i\overline{\imath}$), and quark-antiquark annihilation
($j\overline{\jmath}\to i\overline{\imath}$), where $i$ and $j$ represent
quark species.
We use the thermal quark masses, $m_i^2(T)\simeq m_i^2 + (2g^2/9)T^2$,
in all the rates. At small mass ($m_i(T)<2T$),
the production is largely dominated by the thermal gluon decay channel. We
obtain numerical and analytic solutions of one-dimensional
hydrodynamic expansion of an
initially pure glue plasma. Our results show that even in a quite
optimistic scenario, all quarks are far from chemical equilibrium
throughout the expansion.
Thermal production of light quarks ($u$, $d$ and $s$) is nearly
independent of species.
Heavy quark ($c$ and $b$) production is quite independent of the
transition temperature
and could serve as a very good probe of the initial temperature.
Thermal quark production measurements could also be used to determine
the gluon damping rate, or equivalently the magnetic mass.

\vfill
\begin{center}
{\em Submitted to Physical Review D}
\end{center}
\vfill
CERN-TH.7038/93\\
November 1993
\vspace*{10mm}
\footnoterule
\vspace*{3pt}
\noindent$^*$On leave of absence from  L.A.P.P., BP110, F-74941
Annecy-le-Vieux Cedex, France. Internet: taltherr@vxcern.cern.ch.\\
$^{\dag}$Current address:
Physics Department, Kent State University, Kent, OH 44242 USA.
Internet: seibert@scorpio.kent.edu.\\
\newpage\setcounter{page}{1} \pagestyle{plain}
     \setlength{\parindent}{12pt}

\section{Introduction}

Recently, there has been much interest in calculating the production rates
for both massless and massive quarks in ultra-relativistic nuclear
collisions [\ref{rmueller}-\ref{rcharm}].  Unfortunately, there have been
no rigorous thermal field theory calculations to compare to approximate
results, probably because the ``naive'' production rates diverge for
massless quarks.  The production rates can be regularized by summing over
hard thermal loops [\ref{rKW}, \ref{rW}], following the treatment of
Braaten and Pisarski [\ref{rBP}].  This is a difficult calculation,
and has not been done so far.

The closest approach to a full calculation has been to use the thermal quark
masses for massless quarks to regulate the divergent matrix elements
[\ref{rmueller}].  However, the thermal corrections for massive quarks are
not included, and these can be important when $m \sim T$.
We also take into account the thermal gluon decay process [\ref{rAS}],
which was never considered in previous works.
In this way, we
can see if the quark production rates in non-thermal QCD simulations of
collisions are similar to those in thermally-equilibrated quark-gluon
matter at equivalent temperatures.

In this paper, we calculate thermal quark production from
thermally-equilibrated quark-gluon matter in ultra-relativistic
nuclear collisions.  Our initial state is a high-temperature gluon plasma
(GP), with gluons in thermal and chemical equilibrium but no quarks or
antiquarks.  Produced quarks are assumed to be in thermal equilibrium, as
thermal equilibration times are short, but not in chemical equilibrium.
Production of low-mass ($m_i<\ 3.5T$) quarks is dominated by thermal gluon
decay ($g \rightarrow i\overline{\imath}$), while gluon fusion ($gg
\rightarrow i\overline{\imath}$) is the most important production
mechanism for high mass ($m_i >\ 3.5T$) quarks.
We also include quark-antiquark annihilation ($j\overline{\jmath}
\rightarrow i\overline{\imath}$), where $i$ and $j$ represent different
quark species, although this process is shown to be unimportant to our
analysis.

In section~\ref{sqpr}, we calculate the rates for the processes that we
include in our simulation.  In section~\ref{sqpianc}, we describe our
simulation and calculate effects of the rates on quark chemical
equilibration and thermal quark production.
We summarize our results in section~\ref{ss}.

\section{Quark production rates} \label{sqpr}

In this section we calculate the rates for the various processes.
We begin with the production rate from quark-antiquark collisions,
$R_{j\overline{\jmath} \rightarrow i\overline{\imath}}$.
We then calculate the rates from gluon fusion and thermal gluon decay,
$R_{gg \rightarrow i\overline{\imath}}$ and $R_{g \rightarrow
i\overline{\imath}}$ respectively.

For most of the collision, the quark densities are below their equilibrium
values, so the quark chemical potentials $\mu_i, \mu_j < 0$.  In the absence
of baryons, the antiquark densities are equal to the quark densities, so
we also have $\mu_{\overline{\imath}}=\mu_i<0$.
As the initial densities are near zero, the
initial values $-\mu_i, -\mu_j \gg T$, in which limit the rate
$R_{j\overline{\jmath}\rightarrow i\overline{\imath}}$ can nearly be
expressed in closed form.  The exact rate near equilibrium is not so
important for our results, so we use these far-from-equilibrium rates
throughout this paper.

There does not yet exist any consistent formalism which is able to
deal with systems out of chemical equilibrium.
In particular, merely substituting non-equilibrium quark and antiquark
distributions for equilibrium Fermi-Dirac distributions is certainly
flawed beyond the one-loop level.
Indeed, in chemical equilibrium, the quark chemical
potential is {\em opposite} to the antiquark chemical potential.  This
guarantees the micro-reversibility relations, which are at the basis of a
reliable perturbative expansion [\ref{KLNT}]. In the present work, we use
non-equilibrium densities (with the same chemical potential for quarks
and antiquarks) at the tree-level only.

Even though quarks are not in chemical equilibrium, they thermalize very
fast with the gluons.  The gluon correction to the thermal
quark mass is
\begin{equation}
m_i^2 (T) \simeq m_i^2 + \frac {2g^2} {9} T^2
,\end{equation}
in the high-frequency limit, which we need for most of our calculations.
As there are very few quarks in the plasma, we can simply drop the thermal
quark contribution to the thermal quark mass. This amounts to using
the same approximation as in the rate calculations,
$-\mu_i,-\mu_j \gg T$.

As $g \simeq 2$, this gives us minimum masses of approximately $T$.
These minimum masses are large enough that we can use the high-mass limits
of quark production rates in our simulation.

\subsection{Quark-antiquark collisions}

We begin by calculating the rate for the process $j\overline{\jmath}
\rightarrow i\overline{\imath}$ [\ref{rMMS}]:
\begin{eqnarray}
R_{j\overline{\jmath} \rightarrow i\overline{\imath}} = \int
\frac {d^3p_j} {(2\pi)^3 2 E_j} \frac {d^3p_{\overline{\jmath}}}
{(2\pi)^3 2 E_{\overline{\jmath}}} \frac {d^3p_i} {(2\pi)^3 2 E_i}
\frac {d^3p_{\overline{\imath}}} {(2\pi)^3 2 E_{\overline{\imath}}}
(2\pi)^4 \delta^4 (p_j^{\mu} + p_{\overline{\jmath}}^{\mu}
- p_i^{\mu} - p_{\overline{\imath}}^{\mu})
\nonumber \\ \times \sum \left| {\cal M}_{j\overline{\jmath} \rightarrow
i\overline{\imath}} \right|^2 f_{\mbox{FD}}(E_j)
f_{\mbox{FD}}(E_{\overline{\jmath}})
\left( 1-f_{\mbox{FD}}(E_i) \right)
\left( 1-f_{\mbox{FD}}(E_{\overline{\imath}}) \right),
\label{eRqQ0}
\end{eqnarray}
where $\left| {\cal M} \right|^2$ is summed over initial and final spins
and colors, and
\begin{equation}
f_{\mbox{FD}}(E_{\overline{\imath}}) = f_{\mbox{FD}}(E_i) =
\frac {1} {e^{(E_i-\mu_i)/T}+1}.
\end{equation}
In the limit $-\mu_i, -\mu_j \gg T$, eq.~(\ref{eRqQ0})
reduces to
\begin{eqnarray}
R_{j\overline{\jmath} \rightarrow i\overline{\imath}} = e^{2\mu_j/T} \int
\frac {d^3p_j} {(2\pi)^3 2 E_j} \frac {d^3p_{\overline{\jmath}}}
{(2\pi)^3 2 E_{\overline{\jmath}}} \frac {d^3p_i} {(2\pi)^3 2 E_i}
\frac {d^3p_{\overline{\imath}}} {(2\pi)^3 2 E_{\overline{\imath}}}
\nonumber \\ \times (2\pi)^4 \delta^4 (p_j^{\mu} + p_{\overline{\jmath}}^{\mu}
- p_i^{\mu} - p_{\overline{\imath}}^{\mu}) \sum \left|
{\cal M}_{j\overline{\jmath} \rightarrow i\overline{\imath}} \right|^2
e^{-(E_j+E_{\overline{\jmath}})/T}. \label{eRqQ1}
\end{eqnarray}
Following Ref.~[\ref{rMMS}], we evaluate all but 5 of the integrals
trivially, obtaining
\begin{equation}
R_{j\overline{\jmath} \rightarrow i\overline{\imath}} = \frac {e^{2\mu_j/T}}
{8 (2\pi)^6} \int_{2M}^{\infty}  \!\!\! d\omega e^{-\omega/T}
\int_0^{(\omega^2-4M^2)^{1/2}} \!\!\! dq \int_{-P_*}^{P_*} \!\!\! dP_0
\int_{-p_*}^{p_*} \!\!\! dp_0 \int_0^{2\pi} \!\!\! d\phi \sum \left|
{\cal M}_{j\overline{\jmath} \rightarrow i\overline{\imath}} \right|^2\!,
\label{eRqQ2} \end{equation}
where ${\cal M}$ is evaluated with
\begin{eqnarray}
s &=& \omega^2-q^2, \\
t &=&  M^2+m^2  -\frac {s} {2} \left( 1 - \frac {4P_0p_0} {q^2} \right)
\nonumber \\
&&+2 \left[ \left( \frac s 4 (1-\frac {4 P_0^2} {q^2}) - M^2 \right)
\left( \frac s 4 (1-\frac {4 p_0^2} {q^2}) - m^2 \right) \right]^{1/2}
\cos\phi, \\
u &=& 2M^2+2m^2-s-t.
\end{eqnarray}
Here $M(m)$ and $P_0(p_0)$ are the mass and half the energy difference
of the heavier (lighter) quark and antiquark, while $\omega$ and $q$ are
the pair energy and momentum in the matter rest frame.  The limits of
integration,
\begin{eqnarray}
P_* &=& \frac {q} {2} \left( 1 - \frac {4M^2} {s} \right)^{1/2},
\label{ePs} \\
p_* &=& \frac {q} {2} \left( 1 - \frac {4m^2} {s} \right)^{1/2},
\label{eps} \end{eqnarray}
arise from the kinematic constraints.

For the process $j\overline{\jmath} \rightarrow i\overline{\imath}$, the
summed squared matrix element is
\begin{equation}
\sum \left| {\cal M}_{j\overline{\jmath}\rightarrow i\overline{\imath}}
\right|^2 = 256 \pi^2 \alpha_S^2
\frac {(M^2+m^2-t)^2 + (M^2+m^2-u)^2 + 2(M^2+m^2)s} {s^2}.
\label{eSMqQ2} \end{equation}
Inserting the summed squared matrix element (\ref{eSMqQ2}) in the
rate~(\ref{eRqQ2}) and evaluating the integrals over $\phi$, $p_0$ and
$P_0$, we obtain
\begin{eqnarray}
R_{j\overline{\jmath} \rightarrow i\overline{\imath}} &=&
\frac {2\alpha_S^2} {3\pi^3} e^{2\mu_j/T}
\int_{2M}^{\infty} \! d\omega e^{-\omega/T}
\int_0^{(\omega^2-4M^2)^{1/2}} \! dq q^2 \nonumber \\
&&\times\left[ \left( 1- \frac {4M^2} {s} \right)
\left( 1- \frac {4m^2} {s} \right) \right]^{1/2}
\left\{ 1 + \frac {2(M^2+m^2)} {s} + \frac {4M^2m^2} {s^2} \right\}.
\label{eRqQ3} \end{eqnarray}
In general, the remaining two integrals must be evaluated numerically.
The rate can be evaluated analytically when $M \ll T$, where
\begin{equation}
R_{j\overline{\jmath} \rightarrow i\overline{\imath}} =
\frac {4} {3\pi^3} \alpha_S^2 T^4 e^{2\mu_j/T}
\left\{ 1 + {\cal O} \left[ M/T \right] \right\},
\label{jjiiT}\end{equation}
and when $M \gg T$, where
\begin{equation}
R_{j\overline{\jmath} \rightarrow i\overline{\imath}} =
\frac {1} {\pi^2} \alpha_S^2 MT^3 e^{2(\mu_j-M)/T}
\left( 1- \frac {3m^4} {4M^4} - \frac {m^6} {4M^6} \right)^{1/2}
\left\{ 1 + {\cal O} \left[ T/M \right] \right\}.
\label{jjiiM}\end{equation}

The two limits are almost the same if evaluated with a mass $M \sim T$; the
ratio of the two rates is $3\pi M/4T$ if the exponential is omitted and the
lighter mass is set to zero.  We use only the result for the heavy quark
limit in our simulation, as the difference in the rates for the light quarks
is then not very large if we give them thermal masses.  Also, the total
rate in this low-mass region is dominated by the gluon decay process.

In the limit of low quark density, the forward and reverse rates differ
only in the chemical potential for the initial-state quarks, so we find
\begin{equation}
\delta R_{j\overline{\jmath} \rightarrow i\overline{\imath}} =
\frac {1} {\pi^2} \alpha_S^2 MT^3 e^{-2M/T}
\left[ 1- \frac {3m^4} {4M^4} - \frac {m^6} {4M^6} \right]^{1/2}
\left( e^{2\mu_j/T} - e^{2\mu_i/T} \right). \label{eqR}
\end{equation}
We use the convention throughout this paper that
\begin{equation}
\delta R_{X \rightarrow Y} = R_{X \rightarrow Y} - R_{Y \rightarrow X}.
\end{equation}

\subsection{Gluon fusion}

The rate $R_{gg \rightarrow i\overline{\imath}}$ has not yet been properly
calculated in perturbative thermal QCD.  Naive calculations diverge for
massless quarks, so the rates must be regulated by including hard thermal
loops.  This has not yet been done; only the dominant contribution for small
$g$ has been calculated [\ref{rAS}].
We again approximate the full calculation by using
the low-density thermal quark masses to regulate the production rates.

The rate for the process $gg \rightarrow i\overline{\imath}$ is [\ref{rMMS}]
\begin{eqnarray}
R_{gg \rightarrow i\overline{\imath}} = \frac {1} {2} \int
\frac {d^3p_1} {(2\pi)^3 2 E_1} \frac {d^3p_2} {(2\pi)^3 2 E_2}
\frac {d^3p_i} {(2\pi)^3 2 E_i} \frac {d^3p_{\overline{\imath}}}
{(2\pi)^3 2 E_{\overline{\imath}}} (2\pi)^4 \delta^4 (p_1^{\mu} + p_2^{\mu}
- p_i^{\mu} - p_{\overline{\imath}}^{\mu}) \nonumber \\
\times \sum \left| {\cal M}_{gg \rightarrow i\overline{\imath}}
\right|^2 f_{\mbox{BE}}(E_1) f_{\mbox{BE}}(E_2),
\end{eqnarray}
where the subscripts 1 and 2 refer to the initial-state gluons and
\begin{equation}
f_{\mbox{BE}} (E) = \frac {1} {e^{E/T} -1}.
\end{equation}
Here we have assumed that the heavy quark density is zero, so that
there is no Pauli-blocking of final states.  We proceed in the same manner
as in our calculation of $R_{j\overline{\jmath} \rightarrow
i\overline{\imath}}$, obtaining
\begin{eqnarray}
R_{gg \rightarrow i\overline{\imath}} &=& \frac {1} {16 (2\pi)^6}
\int_{2M_i}^{\infty} d\omega \int_0^{(\omega^2-4M_i^2)^{1/2}} dq
\int_{-p_*}^{p_*} dp_0 \int_{-p_*'}^{p_*'} dp_0'
\nonumber \\ &&\times
\int_0^{2\pi} d\phi
\sum \left| {\cal M}_{gg \rightarrow i\overline{\imath}} \right|^2
f_{\mbox{BE}} \left( \frac {\omega} {2} +p_0 \right)
f_{\mbox{BE}} \left( \frac {\omega} {2} -p_0 \right),
\end{eqnarray}
where $\omega$, $q$ and $\phi$ are defined as before,
\begin{eqnarray}
p_0 &=& E_1 - E_2, \\
p_0' &=& E_i - E_{\overline{\imath}},
\end{eqnarray}
and the limits of integration are
\begin{eqnarray}
p_* &=& \frac q 2, \\
p_*'&=& \frac {q} {2} \left( 1 - \frac {4M_i^2} {s} \right)^{1/2}.
\end{eqnarray}

Up to this point, we have made no approximations (except for ignoring the
final-state Pauli blocking).  Now, we want to perform the calculation
 analytically for the case $M_i \gg T$.  In this case, we again define
$\omega_* = \omega-2M_i$, obtaining
\begin{eqnarray}
R_{gg \rightarrow i\overline{\imath}} &=& \frac {1} {16 (2\pi)^6}
\int_{0}^{\infty} d\omega_* \int_0^{(4M_i\omega_*+\omega_*^2)^{1/2}} dq
\int_{-p_*}^{p_*} dp_0 \int_{-p_*'}^{p_*'} dp_0'
\nonumber \\ &&\times
\int_0^{2\pi} d\phi
\sum \left| {\cal M}_{gg \rightarrow i\overline{\imath}} \right|^2
f_{\mbox{BE}} \left( M_i + \frac {\omega_*} {2} +p_0 \right)
f_{\mbox{BE}} \left( M_i + \frac {\omega_*} {2} -p_0 \right). \label{eRg1}
\end{eqnarray}

It is clear from eq.~(\ref{eRg1}) that $\omega_* \sim T$, so that $p_0 \sim
q \sim (4M_i\omega_*)^{1/2} \sim (M_iT)^{1/2} \ll M_i$.  Thus, the argument of
$f_{\mbox{BE}}$ is always large, so we can replace the Bose distributions by
exponentials.  Similarly, we can show that
\begin{eqnarray}
s &=& 4M_i^2 \left\{ 1+ {\cal O} \left[ T/M_i \right] \right\}, \\
t = u &=& -M_i^2 \left\{ 1+ {\cal O} \left[ (T/M_i)^{1/2} \right] \right\}.
\end{eqnarray}
We can then easily evaluate the integrals to obtain
\begin{equation}
R_{gg \rightarrow i\overline{\imath}} =
\frac {M_iT^3 e^{-2M_i/T}} {512 \pi^4}
\sum \left| {\cal M}_{gg \rightarrow i\overline{\imath}} \right|^2
\left\{ 1 + {\cal O} \left[ T/M_i \right] \right\},
\label{eRg2} \end{equation}
where ${\cal M}$ is evaluated at $s=4M_i^2, t=u=-M_i^2$.  Note that the
fractional correction in $R$ is of higher order in $T/M_i$ than the
fractional correction in $t$ or $u$; this is because ${\cal M}$ is
invariant under the transformation $t \leftrightarrow u$, and
because $s+t+u=2M_i^2$ the correction to $R$ of
${\cal O} \left[ (T/M_i)^{1/2} \right]$ vanishes.

We do not evaluate the summed matrix elements here, as this is done in
Refs.~[\ref{Com}] and [\ref{rMMS}]:
\begin{equation}
\sum \left| {\cal M}_{gg \rightarrow Q\overline{Q}} \right|^2
= \frac {7 \times 256 \pi^2} {3} \alpha_S^2.
\end{equation}
Inserting this into eq.~(\ref{eRg2}), we find
\begin{equation}
R_{gg \rightarrow i\overline{\imath}} =
\frac {7} {6 \pi^2} \alpha_S^2 M_iT^3 e^{-2M_i/T}
\left\{ 1 + {\cal O} \left[ T/M_i \right] \right\}.
\label{ggiiM}\end{equation}
Again, the forward and reverse reactions differ only in chemical potential,
so we obtain
\begin{equation}
\delta R_{gg \rightarrow i\overline{\imath}} =
\frac {7} {6 \pi^2} \alpha_S^2 M_iT^3 e^{-2M_i/T}
\left( 1 - e^{2\mu_i/T} \right). \label{egR}
\end{equation}

\subsection{Thermal gluon decay}

In a previous work [\ref{rAS}], we found that the thermal gluon decay
into quark-antiquark pairs is the leading contribution in the
perturbative expansion, at least for massless quarks. As it is also
relevant for massive quarks, we include it in our numerical simulation.

A diagrammatic analysis reveals that one of the diagrams considered
previously in the gluon fusion process, namely the $s$-channel, is in
fact included in the thermal gluon decay set of diagrams.
There is no over-counting, though, as we do not integrate the same phase
space region in the two cases.

We evaluate the integrals for $R_{g \rightarrow i\overline{\imath}}$ in the
limit $M_i/T \gg 1, g$.  We begin with the expression for the quark
production rate from transverse gluon decay,
\begin{equation}
R^T_{g \rightarrow i\overline{\imath}} = \frac {4g^2} {9\pi^4}
\int_{2M_i}^{\infty} d\omega\ \omega f_{\mbox{BE}} (\omega)
\int_0^{\sqrt{\omega^2-4M_i^2}}
dq\ q^2 \frac { \gamma_g \left( \omega^2-q^2+2M_i^2 \right)
\left[ 1- \frac {4M_i^2} {\omega^2-q^2} \right]^{1/2}}
{\left[ \omega^2 -q^2 -\mbox{Re}\Pi_T(Q) \right]^2
+4 \omega^2 \gamma_g^2}, \label{elm}
\end{equation}
where $\gamma_g$ is the gluon damping rate and $\Pi_T(Q)$ is the
transverse projection of the polarization tensor (the square of the
thermal mass for transverse gluons).
In the region where the integrand is large,
\begin{equation}
\gamma_g={3\over 2}\alpha_S T \ln{1\over \alpha_S} + O(\alpha_S T)
\qquad\mbox{and} \qquad
\mbox{Re}\Pi_T(Q) = {3\over 2}\omega_0^2 = 2\pi\alpha_S T^2
.\end{equation}
In the high-mass and small coupling limit, eq.~(\ref{elm}) reduces to
\begin{equation}
R^T_{g \rightarrow i\overline{\imath}} = \frac {2g^2} {9\pi^4}
\int_{2M_i}^{\infty} d\omega\ \omega e^{-\omega/T} \int_{(2M_i)^2}^{\omega^2}
dm_*^2\ (\omega^2-m_*^2)^{1/2} \frac { \gamma_g \left( m_*^2
+2M_i^2 \right) \left[ 1- \frac {4M_i^2} {m_*^2} \right]^{1/2}}
{\left[ m_*^2 -\mbox{Re}\Pi_T(Q) \right]^2
+4 \omega^2 \gamma_g^2}, \label{elm2}
\end{equation}
where $m_*^2=\omega^2-q^2$.  The integral falls off exponentially fast
with increasing $\omega$, so we expand in $\omega_*=\omega-2M_i$ and
$\delta  m_*^2=m_*^2-4M_i^2$:
\begin{equation}
R^T_{g \rightarrow i\overline{\imath}} = \frac {4g^2 M_i^2 \gamma_g}
{3\pi^4} e^{-2M_i/T} \int_0^{\infty} d\omega_*\ e^{-\omega_*/T}
\int_0^{4M_i\omega_*} d(\delta m_*^2)\ \frac
{\left[ \delta m_*^2 (4M_i\omega_* -\delta m_*^2) \right]^{1/2}}
{\left[ 4M_i^2 -\mbox{Re}\Pi_T(Q) \right]^2 +16 M_i^2 \gamma_g^2}, \label{elm3}
\end{equation}
times $[1+{\cal O} (T/2M_i)]$.  The $\Pi_T$ and $\gamma_g$ in the denominator
provide correction factors $[1+\Pi_T/2M_i^2]$ and $[1-\gamma_g^2/M_i^2]$
respectively, and these can also be dropped if $M_i$ is large enough, giving
\begin{equation}
R^T_{g \rightarrow i\overline{\imath}} = \frac {4 g^2 \gamma_g}
{3\pi^4} e^{-2M_i/T} \int_0^{\infty} d\omega_*\ e^{-\omega_*/T} \omega_*^2
\int_0^1 dz\ \left[ z(1-z) \right]^{1/2}, \label{elm4}
\end{equation}
where $z=\delta m_*^2/4M_i\omega_*$.
These integrals can be evaluated analytically, giving
\begin{equation}
R^T_{g \rightarrow i\overline{\imath}} = \frac {g^2} {3\pi^3}
\gamma_g T^3 e^{-2M_i/T}. \label{elmf}
\end{equation}

The longitudinal case is the same except for a factor of 1/2:
\begin{equation}
R^L_{g \rightarrow i\overline{\imath}} = \frac {g^2} {6\pi^3}
\gamma_g T^3 e^{-2M_i/T}. \label{elmfl}
\end{equation}
The total net production rate is
\begin{equation}
\delta R_{g \rightarrow i\overline{\imath}} = \frac {2 \alpha_S} {\pi^2}
\gamma_g T^3 e^{-2M_i/T} \left( 1 - e^{2\mu_i/T} \right).
\label{eRQ}
\end{equation}
This rate is significantly enhanced for light quarks, however, so we
use the exact rate in our simulation.
\begin{eqnarray}
\delta R_{g \rightarrow i\overline{\imath}}&=&\frac {4 \alpha_S \gamma_g}
{9\pi^3} \left( 1-e^{2\mu_i/T} \right) \int_{2M_i}^{\infty} d\omega \,
\omega \, f_{\mbox{BE}} (\omega) \int_{4M_i^2}^{\omega^2} ds \,
\left( s+2M_i^2 \right) \nonumber \\   &&
\times \left[ \left( \omega^2-s \right)
\left( 1- \frac {4M_i^2} {s} \right) \right]^{1/2}
\left[ \frac {2} {\left( s -3\omega_0^2/2 \right)^2
+4 \omega^2 \gamma_g^2} + \frac {1} {s^2 +4 \omega^2 \gamma_g^2} \right].
\label{edR} \end{eqnarray}

\subsection{Comparison of the rates}

We can now numerically compare the different rates that we have obtained
and comment on our different approximations.
In Fig.~1, we show the quark production rate as a function of the quark
mass $m$, taking $\mu \rightarrow -\infty$ for outgoing quarks.
The thermal gluon decay dominates for $m<2\ T$ and is about
ten times larger than $R_{j\overline{\jmath}\to i\overline{\imath}}$
and $R_{gg\to i\overline{\imath}}$, at $m\simeq 0.5\ T$.
For the latter two processes, we have used the high mass formulas,
eqs.~(\ref{jjiiM}) (with $\mu_j=0$), and (\ref{ggiiM})
respectively.
As $m_i \rightarrow 0$, the quark-antiquark annihilation rate
tends asymptotically to $4/(3\pi^2)\alpha_S^2\simeq 3\times 10^{-3}$,
according to eq.~(\ref{jjiiT}).
Therefore, it is clear that we can reliably
use the high-mass formulas down to $m\simeq 0.5\ T$.
The thermal masses being of order
$T$, we see that we do not need in our simulation anything other than those
high-mass formulas. It makes irrelevant the only
uncertainty in our calculation regarding the process
$R_{gg\to i\overline{\imath}}$,
which has not been calculated rigorously in the low-mass
region. Note also that we have neglected the thermal gluon mass in
calculating $R_{gg\to i\overline{\imath}}$,
which is justified as this process starts to
dominate only when $m >\ 3.5\ T$, which is typically large compared to the
thermal mass.

For the thermal gluon decay, the asymptotic expressions (\ref{elmf}) and
(\ref{elmfl}) turn out to be close to the exact value only when
$m>2\ T$. We have not been able to make a good fit of the numerical curve
for any values of $g$.
We therefore use the exact expression for the gluon decay, eq.~(\ref{edR}),
in our simulation.

\section{Quark production in a nuclear collision} \label{sqpianc}

Our simulation works as follows.  We assume that at initial proper time
$\tau_0$ the gluons are in thermal and chemical equilibrium at initial
temperature $T_0$ but no quarks are present.  We then perform
boost-invariant one-dimensional (no transverse component) hydrodynamic
expansion, conserving all quark numbers but not gluons, then allow
number-changing processes explicitly via chemical reactions.  All particles
are assumed to be in thermal equilibrium (and the gluons in chemical
equilibrium) throughout the collision.

The evolution equations are
\begin{eqnarray}
\frac {de} {d\tau} &=& -\frac {e+P} {\tau}, \label{ehc1} \\
\frac {dn_i} {d\tau} &=& \frac {-n_i} {\tau} +\delta R_{g \rightarrow
i\overline{\imath}} +\delta R_{gg \rightarrow i\overline{\imath}}
+\sum_j \delta R_{j\overline{\jmath} \rightarrow i\overline{\imath}}.
\label{ehc2} \end{eqnarray}
The first terms of eqs.~(\ref{ehc1}) and (\ref{ehc2}) give the
hydrodynamic evolution, and the remaining three terms of eq.~(\ref{ehc2})
give the chemical evolution of the quarks.

The temperature and chemical potentials are recalculated after each time
step from the densities and chemical potentials, using the thermal
equilibrium conditions
\begin{eqnarray}
n_i (T, \mu_i) = n_{\overline{\imath}} (T, \mu_i) &=& \frac {3} {\pi^2}
\int_0^{\infty} dk \, k^2 \, f_{\mbox{FD}} \left( E_i(k) \right),
\label{eneq} \\
e (T, \{\mu_i\}) &=& \frac {8 \pi^2} {15} T^4 + \sum_i \frac {6} {\pi^2}
\int_0^{\infty} dk \, k^2 \, E_i(k) \, f_{\mbox{FD}} \left( E_i(k) \right), \\
P (T, \{\mu_i\}) &=& \frac {8 \pi^2} {45} T^4 + \sum_i \frac {2} {\pi^2}
\int_0^{\infty} \frac {dk \, k^4}{E_i(k)} \, f_{\mbox{FD}} \left( E_i(k)
\right).
\end{eqnarray}
The first contributions to the energy density and the pressure are from the
gluons.  All of the above integrals are evaluated numerically to better than
1\% accuracy.

We use the same running coupling constant as in Ref.~[\ref{rg}]:
\begin{equation}
\alpha_S = \frac {12 \pi} {(33-2n_f) \ln \left[ \overline{Q^2} / \Lambda^2
\right]}. \label{erun}
\end{equation}
Here $n_f$ is the number of light flavors for the given reaction (4 for
$b$ production, 3 for all other reactions), $\overline{Q^2}$ is the mean
momentum transfer, and $\Lambda$ is the renormalization scale.
We use $\Lambda=300$ MeV for our calculations, instead of 400 MeV as in
Ref.~[\ref{rg}]; this gives larger values of $\alpha_S$ and thus higher
reaction rates.
We take the ansatz
\begin{equation}
\overline{Q^2} = m^2(T) + 9T^2,
\end{equation}
which is correct in the high-mass limit and for a Boltzmann distribution
in the low-mass limit. We have checked that this $\alpha_S$-behavior is in
good agreement with recent calculations on the running coupling constant at
finite temperature [\ref{ESW}]. It gives a rather weak $\Lambda$-dependence.

For our first simulation, we use conditions for an Au+Au reaction
at $\sqrt{s}=200$ GeV/nucleon:
\begin{eqnarray}
\tau_0 &=& 0.063\ \mbox{fm}/c, \\
T_0 &=& 950\ \mbox{MeV},
\end{eqnarray}
The initial time is somewhat arbitrary; we picked these numbers because
they give the same mass spectrum for thermal dilepton production as a
more detailed collision simulation [\ref{rSA1}].  For quark masses we
use $m_u=m_d=0$, $m_s = 0.2$, $m_c=1.5$ and $m_b=5$ GeV.
We run until $\tau=11$ fm/c, corresponding to $T \approx 150$ MeV.
We refer to this as our optimistic scenario because the extremely short
equilibration time (and correspondingly high temperature) result in large
values of heavy quark production.

In Fig.~2 we show the ratio of density to chemical equilibrium ($\mu=0$)
density for each quark species.  This ratio gives a good measure of the
degree of chemical equilibration for the light quarks ($u$, $d$ and $s$),
as they are produced throughout the evolution.
Note that, because of the thermal mass effects, thermal production of
light quarks is nearly independent of species.
Heavy quarks ($c$ and $b$)
are produced only in the early stages, when their densities are far below
equilibrium values.  In the later stages their densities actually exceed
the equilibrium values, but at this point they are no longer interacting
chemically so they are not equilibrated (they pass the equilibrium values
because the equilibrium densities fall exponentially fast with increasing
$\tau$).

We see from Fig.~2 that all quarks are far from chemical equilibrium
throughout the collision, in agreement with the conclusions of Geiger and
Kapusta~[\ref{rgkce}].  However, our results differ somewhat from
theirs.  The equilibration times are longer, so that ratios of density
to equilibrium density rise more slowly at first.  Also, the ratios of density
to equilibrium density are always increasing throughout our simulation;
this is because all quark densities are less than their equilibrium
values except for the heavy quarks after freezeout.

In Figs.~3 and 4, we show the quark (or antiquark) production per unit
rapidity at $y=0$ for each species as a function of $T_c$, using
respectively our optimistic scenario and a more standard scenario with
the same entropy density:
\begin{eqnarray}
\tau_0 &=& 0.2\ \mbox{fm}/c, \\
T_0 &=& 650\ \mbox{MeV}.
\end{eqnarray}
The rapidity density is simply related to the density at $T_c$:
\begin{equation}
\left. dN_i/dy \right|_{y=0} = \left. A \tau n_i \right|_{T=T_c}.
\end{equation}
where $A$ is the cross-sectional area of the plasma.  We use $A=150$
fm$^2$ for a central Au+Au collision.

These results can be understood analytically.  For
$M \gg T_0$, the heavy quark approximations are clearly justified.
The dominant production mechanism is then gluon fusion, as $R_{gg
\rightarrow Q\overline{Q}}/R_{g \rightarrow Q\overline{Q}} \propto
M/T$.  Quark collisions are unimportant, as all quark densities are
small throughout the collision.

Because the quark densities are small, the thermal behaviour is dominated
by gluons, so that to a good approximation
\begin{equation}
\tau T^3 = \tau_0 T_0^3,
\end{equation}
as expected from entropy conservation.  We can then estimate the quark
densities:
\begin{equation}
n_Q^{gg} (\tau) = \frac {1} {\tau} \int_{\tau_0}^{\tau} d\tau' \, \tau' \,
R_{gg \rightarrow i\overline\imath} (\tau')
= \frac {7 \alpha_S^2} {2 \pi^2} M \tau_0 T_0^3 T^3 \int_{T}^{T_0}
\frac {dT'} {T^{\prime 4}} e^{-2M/T'}.
\end{equation}
We then obtain the contribution to the rapidity density from thermal
gluon-gluon collisions in the plasma,
\begin{equation}
dN_Q^{gg}/dy = \frac {7 \alpha_S^2} {4 \pi^2} A \tau_0^2 T_0^4
e^{-2M/T_0} \left[ 1 +\frac {T_0} {M} +\frac {T_0^2} {2M^2}
-\frac {T_0^2} {T_c^2} e^{2M/T_0-2M/T_c} \left( 1 +\frac {T_c} {M}
+\frac {T_c^2} {2M^2} \right) \right].
\end{equation}

We can make a similar calculation of the contribution from thermal gluon
decay, obtaining
\begin{equation}
dN_Q^g/dy = \frac {9 \alpha_S^2 \ln \frac {1} {\alpha_S}} {2 \pi^2}
A \tau_0^2 \frac {T_0^5} {M} e^{-2M/T_0} \left[ 1 +\frac {T_0} {2M}
-\frac {T_0} {T_c} e^{2M/T_0-2M/T_c}
\left( 1 +\frac {T_c} {2M} \right) \right].
\end{equation}
When $2M/T_c - 2M/T_0 \gg 1$, the heavy quark production in the plasma is
independent of $T_c$:
\begin{equation}
dN_Q/dy = \frac {7 \alpha_S^2} {4 \pi^2} A \tau_0^2 T_0^4 e^{-2M/T_0}
\left[ 1 +\left(1+ \frac {18} {7} \ln \frac {1} {\alpha_S} \right)
\frac {T_0} {M} + \frac {1} {2} \left(1+\frac {18} {7}
\ln \frac {1} {\alpha_S} \right) \frac {T_0^2} {M^2} \right].
\label{ehqp} \end{equation}
This is because heavy quark production occurs only at the earliest
(and hottest) part of the collision.

In Fig.~5, we compare $c$ and $b$ quark production in our
simulation with the analytic formula for heavy quarks (\ref{ehqp}).
We calculate $\alpha_S$ from eq.~(\ref{erun}) with $T=T_0$, and use
this value to calculate $\gamma_g$ and the thermal mass; this works
well because most of the heavy quark production occurs near $T_0$.
The analytic result is a bit high for the $b$ quarks because the
quark production lowers the temperature faster than for a pure glue
system.

The situation for massless quarks is somewhat different.  Here the
dominant production mechanism is decay of thermal gluons.  We thus
find
\begin{eqnarray}
R_{g \rightarrow q\overline{q}} &=& KT^4, \\
K &=& \frac {8 \alpha_S \gamma_g}
{9 \pi^3} \int_{2\nu}^{\infty} d\omega \, \omega \,
f_{BE} (\omega) \int_{4\nu^2}^{\omega^2} ds \, \left( s+ 2\nu^2 \right)
\nonumber \\ && \times \left[ \left( \omega^2-s \right)
\left( 1- \frac {4\nu^2} {s} \right) \right]^{1/2}
\left\{ \frac {2} {(s-2\pi\alpha_S)^2 + 4 \omega^2 \gamma_g^2}
+ \frac {1} {s^2 + 4 \omega^2 \gamma_g^2} \right\}.
\end{eqnarray}
Here $\nu$ is the ratio of the thermal quark mass to $T$:
\begin{equation}
\nu = \left( \frac {8\pi \alpha_S} {9} \right)^{1/2}.
\end{equation}
We then obtain
\begin{equation}
dN_q^g/dy = \frac {3} {2} K A \tau_0^2 T_0^6 T_c^{-2} \left( 1-
\frac {T_c^2} {T_0^2} \right).
\end{equation}

Similarly, we find
\begin{equation}
dN_q^{gg}/dy = \frac {7} {4\pi^2} \alpha_S^2 \nu e^{-2\nu} A \tau_0^2
T_0^6 T_c^{-2} \left( 1- \frac {T_c^2} {T_0^2} \right). \label{elqp}
\end{equation}
Note that, for massless quarks, when $T_0 \gg T_c$ we find that
\begin{equation}
dN_q/dy = \left( \frac {3} {2} K + \frac {7} {4\pi^2} \alpha_S^2 \nu
e^{-2\nu} \right) A \tau_0^2 T_0^6 T_c^{-2}.
\end{equation}
Light quark production in the plasma is not independent of $T_c$.
The difference between analytic and simulated results can be seen in
Fig.~6. The agreement is not so good as in the heavy quark case but still
remains within a factor of two. The
other main difference from the heavy quark production is that the best
value of $\alpha_S$ to use is not given by the value at $T_0$, as the
light quark production is not dominated so much by the early stages.
Of course, this quantity is also relatively meaningless for the $u$ and
$d$ quarks, as these are copiously produced in normal hadronic (confined)
matter at temperatures near $T_c$.  However, the $s$ quarks are not so
easily produced in confined matter, so their production in the plasma is
of interest.  Also, the analytic result serves as a useful check of the
simulation.

Let us finally list some remarks:

-- The sensitivity to the choice of the renormalization scale $\Lambda$ is
rather weak. Our results vary less than 30\% for $\Lambda=300-400$ MeV
At the beginning of the expansion, $\alpha_S$ is always small and
perturbative QCD applies, so our calculation of heavy quark
production is reliable.
Light quarks are still being produced at the end of the plasma phase
(near $T_c$), where large values of $\alpha_S$ can lead
to significant quark-antiquark production and annihilation.
The subsequent hadronization also introduces large uncertainties in light
quark production, especially for the $u$ and $d$ flavors.

-- An exciting feature is that the gluon decay process
is responsible for much of the heavy quark production.
The simulations show that, with $T_0=950$ MeV, it accounts for
70\% of charm production and 50\% of bottom production.  If $T_0=650$ MeV,
it still represents 60\% of charms and 40\% of bottoms which are produced
during the expansion.
These numbers agree well with the analytic estimate (\ref{ehqp}).
Therefore, it seems very realistic to measure the gluon damping rate, or
equivalently the magnetic mass, at future heavy-ion colliders if a plasma
with sufficiently high temperature can be formed.

\section{Conclusions} \label{ss}

We have calculated thermal quark production in ultra-relativistic heavy ion
collisions. The following processes have been taken into account:
thermal gluon decay ($g\to i\overline{\imath}$), gluon fusion
($gg\to i\overline{\imath}$), and quark-antiquark annihilation
($j\overline{\jmath}\to i\overline{\imath}$).
We have used the thermal quark masses
in all the rates.  Quark production is dominated by the
gluon decay channel for small quark mass, and by gluon fusion for high mass.

We have performed a one-dimensional hydrodynamic expansion of a plasma,
initially free of quarks. Our results show that even in a quite
optimistic scenario, that is for a simulation with initial time
$\tau_0=0.063$ fm/c and an initial temperature $T_0=950$ MeV, the light
quarks ($u$, $d$ and $s$) do not reach equilibrium at the end of the expansion.
Furthermore, due to thermal mass effects, light quark ($u$, $d$ and $s$)
production is nearly independent of species.
Heavy quark ($c$ and $b$) production is
quite independent of the transition temperature and could serve as a very
good probe of the initial temperature.
Thermal quark production could also be used to determine the gluon damping
rate, provided the initial heavy quark density created in parton collisions
(before the formation of the plasma) is not too large.  This is equivalent
to determining the magnetic mass.  All these numerical results have been
checked to be in very good agreement with simple analytic estimates, at
least for the heavy quarks.  For light quarks, the agreement between
analytic and numerical results is within a factor of two.

Obviously, these results should have deep implications for other signals,
as thermal photon and dilepton emission, for which quark chemical equilibrium
is usually assumed [\ref{KLNT},\ref{Vesa}].
At the moment, it is not clear whether the
relatively high initial temperature can counterbalance the effect of having
four to five times less quarks than in a complete equilibrium picture.

Finally, the large rates of heavy quark production seen in the parton cascade
model [\ref{rg}] do not appear in our simulation. This indicates that they
are probably a result of high-energy parton collisions, and not from any
quark-gluon plasma that is included (implicitly) in the parton cascade model.

\ulsect{Acknowledgements}

T.A. would like to thank Ivan Dadi\'c for a pertinent remark on the
micro-reversibility conditions.
This material is based upon work supported by the North Atlantic Treaty
Organization under a Grant awarded in 1991.

\vfill \eject

\ulsect{References}

\begin{list}{\arabic{enumi}.\hfill}{\setlength{\topsep}{0pt}
\setlength{\partopsep}{0pt} \setlength{\itemsep}{0pt}
\setlength{\parsep}{0pt} \setlength{\leftmargin}{\labelwidth}
\setlength{\rightmargin}{0pt} \setlength{\listparindent}{0pt}
\setlength{\itemindent}{0pt} \setlength{\labelsep}{0pt}
\usecounter{enumi}}

\item T. Bir\'o, E. van Doorn, B. M\"uller, M. Thoma and X. Wang,
Phys.\ Rev.\ C {\bf 48}, 1275 (1993). \label{rmueller}

\item K. Geiger, Phys.\ Rev.\ D {\bf 46}, 4965 (1992). \label{rg}

\item E. Shuryak, Phys.\ Rev.\ Lett.\ {\bf 68}, 3270 (1992). \label{rs}

\item T. Matsui, B. Svetitsky and L. McLerran, Phys.\ Rev.\ D {\bf 34},
783 (1986). \label{rMMS}

\item J. Rafelski and B. M\"uller, Phys.\ Rev.\ Lett.\ {\bf 48}, 1066
(1982). \label{rstrange}

\item A. Shor, Phys. Lett. B {\bf 215}, 375 (1988). \label{rcharm}

\item V. Klimov, Sov.\ J. Nucl.\ Phys.\ {\bf 33}, 934 (1981);
      H.A. Weldon, Phys.\ Rev.\ D {\bf 26}, 1394 (1982).  \label{rKW}

\item H.A. Weldon, Phys.\ Rev.\ D {\bf 26}, 2789 (1982). \label{rW}

\item E. Braaten and R. Pisarski, Nucl.\ Phys.\ {\bf B337}, 569 (1990);
      Nucl.\ Phys.\ {\bf B339}, 310 (1990).\label{rBP}

\item T.~Altherr and D.~Seibert, Phys.\ Lett.\ B {\bf 313}, 149 (1993).
      \label{rAS}

\item J.~Cleymans and I.~Dadi\'c, Z. Phys.\ C {\bf 42}, 133 (1989);
      T.~Altherr, P.~Aurenche and T.~Becherrawy, Nucl.\ Phys.\ {\bf B315},
      436 (1989);
      R.~Baier, B.~Pire and D.~Schiff, Phys.\ Rev.\ D {\bf 38}, 2814 (1988);
      Y.~Gabellini, T.~Grandou and D.~Poizat, Ann.\ Phys.\ (NY)
      {\bf 202}, 436 (1990);
      T.~Altherr and T.~Grandou, Nucl.\ Phys.\ {\bf B402}, 195 (1993).
\label{KLNT}

\item B. Combridge, Nucl.\ Phys.\ {\bf B151}, 429 (1979). \label{Com}

\item M.~A.~van Eijck, C.~R.~Stephens and Ch.~G.~van Weert, preprint
ITFA-93-11 (unpublished). \label{ESW}

\item D. Seibert and T. Altherr, Phys.\ Rev.\ D {\bf 48}, 3386 (1993).
\label{rSA1}

\item K.~Geiger and J.~I.~Kapusta, Phys.\ Rev.\ D {\bf 47}, 4905 (1993).
\label{rgkce}

\item P.~V.~Ruuskanen, Proc.\ of Quark Matter '90, Nucl.\ Phys.\ {\bf A525},
255c (1991). \label{Vesa}
\end{list}
\vfill\eject

\ulsect{Figure captions}

\begin{list}{\arabic{enumi}.\hfill}{\setlength{\topsep}{0pt}
\setlength{\partopsep}{0pt} \setlength{\itemsep}{0pt}
\setlength{\parsep}{0pt} \setlength{\leftmargin}{\labelwidth}
\setlength{\rightmargin}{0pt} \setlength{\listparindent}{0pt}
\setlength{\itemindent}{0pt} \setlength{\labelsep}{0pt}
\usecounter{enumi}}

\item Comparison of quark production rates in the different channels.
We use the complete result for the gluon decay, eq.~(\ref{edR}), and
high-mass formulas for gluon fusion and quark-antiquark annihilation,
eqs.~(\ref{ggiiM}) and (\ref{jjiiM}) respectively,
with $\alpha_s=0.32$.  Outgoing quarks are unthermalized, and incoming
quarks ($u$ and $d$) are in thermal equilibrium ($\mu_u=\mu_d=0$).

\item Degree of chemical equilibration in a central Au+Au collision
with $\sqrt{s}=200$ GeV/nucleon for the optimistic scenario
($\tau_0=0.063$ fm/c, $T_0=950$ MeV).

\item Rapidity densities in a central Au+Au collision with $\sqrt{s}=200$
GeV/nucleon for the optimistic scenario ($\tau_0=0.063$ fm/c, $T_0=950$ MeV).

\item Rapidity densities in a central Au+Au collision with $\sqrt{s}=200$
GeV/nucleon for the standard scenario ($\tau_0=0.2$ fm/c, $T_0=650$ MeV).

\item Comparison of simulation with analytic results for heavy quark
production, eq.~(\ref{ehqp}), for the optimistic scenario
($\tau_0=0.063$ fm/c, $T_0=950$ MeV).

\item Comparison of simulation with analytic results for light quark
production, eq.~(\ref{elqp}), for the optimistic scenario
($\tau_0=0.063$ fm/c, $T_0=950$ MeV).

\end{list}

\end{document}